\documentclass[journal=iecred,manuscript=article]{achemso}
\usepackage[version=3]{mhchem} 
\usepackage[T1]{fontenc}       
\usepackage{graphicx}
\usepackage{amsmath}
\usepackage{pdflscape}
\usepackage{multirow}
\usepackage{multicol}
\usepackage{geometry}
\usepackage{textcomp}
\usepackage{float}

\usepackage{float}

\author{Somasekhara Goud Sontti}

\author{Arnab Atta}
\email{arnab@che.iitkgp.ernet.in}
\phone{+91 (3222) 283910}
\affiliation[mCFD]
{Multiscale Computational Fluid Dynamics (mCFD) Laboratory, Department of Chemical Engineering, Indian Institute of Technology Kharagpur, West Bengal 721302, India}

\title[Taylor bubble in Newtonian and non\textendash Newtonian liquid]
{CFD analysis of Taylor bubble in a co\textendash flow microchannel with Newtonian and non\textendash Newtonian liquid}

\begin{document}
\begin{abstract}

We present a CFD based model to understand the Taylor bubble behavior in Newtonian and non-Newtonian liquids flowing through a confined co\textendash flow microchannel. Systematic investigation is carried out to explore the influence of surface tension, inlet velocities, and apparent viscosity on the bubble length, shape, velocity, and film thickness around the bubble. Aqueous solutions of carboxymethyl cellulose (CMC) with different concentrations are considered as power\textendash law liquid to address the presence of non\textendash Newtonian continuous phase on Taylor bubble. In all cases, bubble length was found to decrease with increasing Capillary number, inlet gas-liquid velocity ratio, and CMC concentration. However, bubble velocity increased due to increasing liquid film thickness around the bubble. At higher Capillary number and inlet velocity ratio, significant changes in bubble shapes are observed. With increasing CMC concentration, bubble formation frequency and velocity increased, but length decreased. 



\end{abstract}






\section{Introduction}

 Two\textendash phase flow in microchannels and microcapillaries has been an area of keen interest over the last few decades, because of its relevance in lab\textendash on\textendash a\textendash chip devices, microreactors \citep{kreutzer-2005}, and several chemical \citep{yao-2015} as well as biological \citep{lee-2015} applications. Considerable attention has been devoted on understanding the hydrodynamics of two\textendash phase flows at the microscale that are important to enhance heat and mass transfer phenomena \citep{kuzmi-2013} as well as to help in chemical process development and intensification \citep{abiev-2012}. Depending on the fluid properties  and microchannel geometry configuration, different flow patterns such as bubble, slug, churn, annular, and stratified flow are observed. Numerous experimental studies have addressed these various flow regimes in microchannels and have presented flow regime maps \citep{tripl-1999, wang-2012}. 

Over the past decades, several observations in T\textendash junction microchannel are reported \citep{van-2007,leclerc-2010}. For analyzing flow pattern transitions in T\textendash junction microchannels, empirical correlations based on fluid properties and the wetting phenomena have been developed by \citet{wael-2006}. Numerical studies on Taylor bubble flow in T\textendash junction using Volume\textendash of\textendash Fluid (VOF) method was carried out by \citet{qian-2006} to understand the influence of inlet configuration, fluid velocity and properties, such as surface tension, liquid viscosity, and wall surface adhesion. However, liquid film thickness was not captured due to poor grid resolution, and different bubble shapes were reported by varying wall adhesion property. Flow\textendash focusing junction is another popular configuration used for bubble generation in microchannel \citep{lua-2016}. \citet{yuu-2007} carried out experiments and Lattice Boltzmann simulations of air\textendash oil two\textendash phase flow in cross and converging shaped microchannels. Different flow regimes (bubbly and slug) and bubble shapes were found depending on Capillary number ($Ca$). \citet{luu-2014} studied bubble formation in a square microchannel using flow\textendash focusing device for higher viscosity liquids and showed that the liquid viscosity strongly influenced bubble formation process, its length, and shape. 

Co\textendash flow configuration is one of the simplest microfluidic geometries that are typically used for Taylor bubble formation, where the dispersed and continuous phases flow in parallel. \citet{salman-2006} studied Taylor bubble formation in capillary tubes using co\textendash flow microchannel and observed different mechanism for bubble formation and coalescence in smaller nozzles with low liquid flow rates. \citet{chen-2007} investigated Taylor bubble formation in a nozzle\textendash tube co\textendash flow configuration using level set method and predicted liquid film thickness around the bubble which was in fair agreement with the experimental observation by \citet{breth-1961}. \citet{goel-2008} and \citet{shaods-2008} also numerically analyzed bubble formation in circular capillaries and described the effect of various parameters, such as superficial velocities, capillary diameter, and wall contact angle. \citet{kawaharaq-2009} examined two\textendash phase flow characteristics in a horizontal circular microchannel by experimental measurement of bubble velocity, void fraction, and pressure drop. To understand the interface morphology of Taylor bubble, \citet{lu-2015} investigated the effect of several working conditions. Similarly, bubble formation in a tapered co\textendash flow geometry \citep{wangg-2015}, flow-focusing microchannel \citep{jia-2016}, and mixing junction \citep{dang-2015} was also reported. Novel methods of bubble formation mechanism and breakup dynamics in different microfluidic devices are comprehensively summarized by \citet{fu-2015b}. For more insight on numerical and experimental understanding of two\textendash phase flow system in microchannel, reader may also refer to other review articles \citep{shao-2009,haase-2016}.

       
\indent Interestingly, most of the reported research are concerned with the bubble formation characteristics and size determination in Newtonian fluids, while several fluids in numerous applications\citep{tsakiroglou2004, nghe2011}, such as emulsion based enhanced oil recovery, drug delivery, food processing, catalytic polymerization reactions, are likely to exhibit non\textendash Newtonian properties. Recently, few experimental studies have depicted the effect of rheological properties on Taylor bubble formation in various microchannel configurations. Using micro\textendash PIV, \citet{fuuu-2012} studied slender bubble formation in flow\textendash focusing device with various concentration of polyacrylamide (PAM) solutions. Gaseous thread width variation with time before pinch\textendash off was investigated and a simple power\textendash law relation was proposed to understand the breakup dynamics. In a T\textendash junction microchannel, \citet{mansour2015} analyzed bubble formation in aqueous solutions of PAM and compared the bubble length, liquid slug length, bubble velocity and frictional pressure drop with those for Newtonian (water) system. They showed significant influence of non-Newtonian rheological properties on the behavior of two\textendash phase flow phenomena. In millimeter-scale T\textendash junction and flow\textendash focusing devices, \citet{labor-2015} studied bubble formation in yield stress fluids. Using a flow\textendash focusing device, \citet{fu-2016} examined droplet breakup dynamics in shear\textendash thinning PAM solution. Incorporating the rheological properties, a simple scaling law was proposed to predict the droplet size. Although, the results showed interesting perspectives on droplet breakup mechanism in non\textendash Newtonian liquids, the necessity of more insight can be evidently be realized for analyzing Taylor bubble characteristics and behavior in non-Newtonian liquids flowing through a co-flow microchannel.

\begin{figure}[h]
	\centering
	\includegraphics[width=0.9\textwidth]{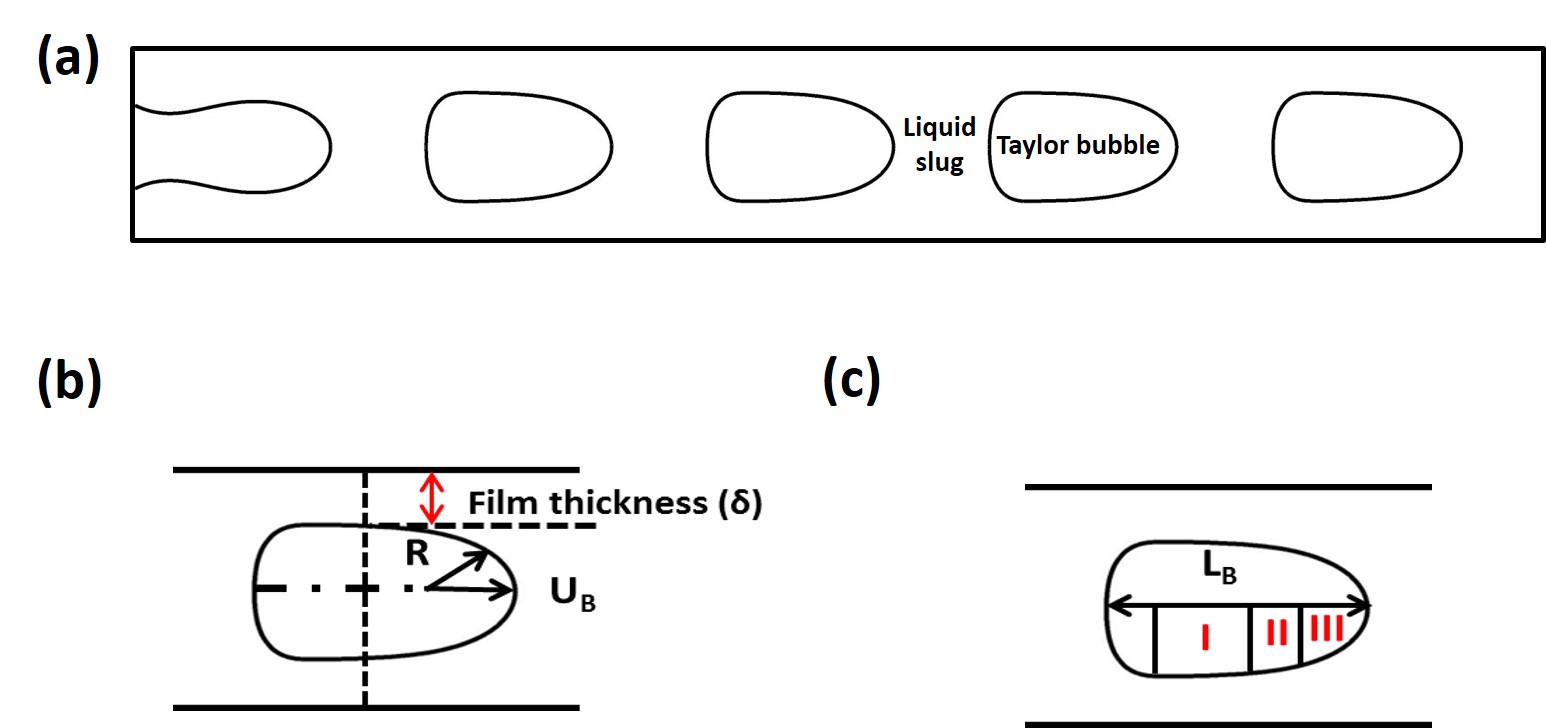}
	\caption{\label{fig:Taylorbubble_model} Schematic representation of (a) Taylor bubble formation in a co\textendash flow geometry, (b) surrounding liquid film thickness, radius of nose (R), and bubble velocity ($U_B$) at the nose, (c) liquid film thickness profiles, Zone I: constant film thickness, Zone II: dynamical meniscus, Zone III: nose region.}
\end{figure}

Moreover, there is considerable argument on capturing and/or identifying the liquid film thickness surrounding a Taylor bubble \citep{Fletcher2016}. Recently, \citet{jia-2016} explored the effects of liquid film thickness and bubble shape characteristics on mass transfer coefficient. They concluded that mass transfer process significantly depended on the liquid film thickness around the Taylor bubble. \citet{boden2016} reported the implication of thin film region during dissolution of $CO_2$ in a square millichannel. In light of the aforementioned discussion, in this article, we present a CFD based model to study the two\textendash phase (gas\textendash liquid) flow in a circular co-flow microchannel with an emphasis on determining the bubble shape, velocity and surrounding liquid film thickness in Newtonian and non-Newtonian liquid phase. A schematic depicting typical Taylor bubble flow in a 2-D co\textendash flow geometry is shown in Fig.~\ref{fig:Taylorbubble_model}a. 

Fig. \ref{fig:Taylorbubble_model}a shows that in a gas\textendash liquid  co\textendash flow system, a Taylor bubble is typically surrounded by a thin liquid film adjacent to the wall which is magnified and illustrated in Fig. \ref{fig:Taylorbubble_model}b. Two consecutive gas bubbles are separated by liquid slugs. In general, Taylor bubble shape can be estimated from three different zones\citep{huerre2014}, as illustrated in Fig. \ref{fig:Taylorbubble_model}c. Depending on the operating conditions and fluid properties, the radius of curvature (R in Fig. \ref{fig:Taylorbubble_model}b) changes in dynamic meniscus (Zone II) and nose region (Zone III) that dictates the bubble shape. In the first half of this article, we report the effect of surface tension and inlet velocities on Taylor bubble and surrounding liquid film thickness ($\delta$) in Newtonian liquid phase using VOF method with  refinement near the solid wall. Thereafter, we extend our investigation for non\textendash Newtonian liquids, that is scarce in the open literature. Aqueous solutions of carboxymethyl cellulose (CMC) with different mass concentration are considered as the non\textendash Newtonian liquid phase which exhibit power\textendash law behavior. Systematic studies are subsequently carried out to realize the influence of CMC concentration on Taylor bubble formation and adjacent liquid film thickness. 

\section{Numerical model}
In general, the physical process of fluid flow are described by a set of conservation equations like mass, momentum and an additional advection equation for a marker function to identify the interface \citep{albad-2013}. To track the interface between two phases, different methods are available namely, Volume of Fluid (VOF) \citep{hirt-1981}, Level Set \citep{suss-1994}, Phase Field \citep{sun-2007} and Lattice Boltzmann \citep{shan-1993l} method. In this study, VOF is utilized and the following set of equations are solved for mass, momentum and volume fraction calculation.

\subsection{Volume of Fluid (VOF) method}

In VOF approach, a single set of conservation equations is solved for immiscible fluids. The governing equations of the VOF formulation for multiphase flows are as follows\cite{ranadebook}:

\subsubsection{Equation of continuity} 

\begin{equation}
\label{eq:mass_eqn}
\frac{\partial  \rho }{\partial t}  +  \nabla .  ( \rho  \vec{ U } ) =0
\end{equation}

\subsubsection{Equation of motion}

\begin{equation}
	\label{eq:mom_eqn}
	\frac{ \partial (\rho \vec{ U })}{ \partial t} + \nabla.( \rho \vec{ U } \vec{ U }) = - \nabla p + \nabla.\tau + \vec{ F}_{SF}
\end{equation}

\begin{equation}
	\label{eq:tau}
	\tau = \eta (\nabla \vec {U} + \nabla { \vec {U} } ^{T}) = \eta \dot{ \gamma }
\end{equation}

\noindent where $\vec{ F}_{SF}$, $\tau$, $p$, $\rho$ and $\eta$ are the volumetric surface tension force, shear stress, pressure, volume-averaged density, and dynamic viscosity, respectively. For non\textendash Newtonian  power\textendash law fluid, apparent viscosity ($\eta$) is expressed as:
\begin{equation}
	\label{eq:nnvis_eqn}
	\eta=K \dot{\gamma }^{n-1} 
\end{equation} 
where $\dot{ \gamma}$, $K$ and $n$ are the  local shear rate, consistency coefficient, and flow index, respectively. For a two\textendash phase system, if the phases are represented by the subscripts and the volume fraction ($\alpha$) of the secondary phase is known, then the density and viscosity in each cell are given by: 
\begin{equation}
\label{eq:frac1_eqn}
\rho =  \alpha _{2}  \rho _{2}  + (1 - \alpha _{2})\rho _{1} 
\end{equation}
\begin{equation}
\label{eq:frac2_eqn}
\eta =  \alpha _{2}   \eta  _{2}  + (1 - \alpha _{2}) \eta  _{1} 
\end{equation}

\subsubsection{Equation of marker function}

In absence of any mass transfer between phases, the interface between the two phases can be tracked by solving the following continuity equation (Eq. \ref{eq:vof_eqn}) for the volume fraction function.
\begin{equation}
\label{eq:vof_eqn} 
\frac{\partial    \alpha _{q}  }{\partial t}  +  ( \vec{ U_q }  .  \bigtriangledown)   \alpha _{q} =0
\end{equation}

where $ q$ denotes either gas or liquid phase. The volume fraction for the primary phase in Eq. \ref{eq:vof_eqn} is then obtained from the following equation:
\begin{equation}
\sum  \alpha _{q} =1
\end{equation} 

\subsection{Surface tension force}
Surface tension effect plays a significant role in droplet deformation, bubble motion and liquid/bubble breakup. The continuum surface force (CSF) model \citep{brack-1992} that has been widely and successfully applied to incorporate surface tension force, is used in this work. Surface tension force ($\vec{ F}_{SF}$) is added to VOF calculation as a source term in the momentum equation. For gas\textendash liquid two\textendash phase flows, the source term in Eq. \ref{eq:mom_eqn} that arises from surface tension can be represented as: 

\begin{equation}
\vec{F} _{SF} = \sigma  \kappa_{n}  \begin{bmatrix} \frac{  \alpha _{1}  \rho _{1}+   \alpha _{2}  \rho _{2}}{ \frac{1}{2}  ( \rho_{1}+  \rho_{2})}   \end{bmatrix} 
\end{equation}

where $\kappa_{n}$ is the radius of  curvature and  $\sigma$ is the surface tension. $\kappa_{n}$ is further defined in terms of the unit normal $ \hat{n} $ as follows \citep{fluent} :

\begin{equation}
\kappa_{n} = -\bigtriangledown  .  \hat{n} = \frac{1}{| n |}  \begin{bmatrix} \big( \frac{n}{ | n |} . \bigtriangledown\big) | n | - \big( \bigtriangledown  . n\big) \end{bmatrix}
\end{equation}

where $\hat{n} = \frac{n}{ | n | } $, and $n= \bigtriangledown   \alpha _{q}$.

\noindent If $ \theta _{ w } $ is the contact angle at the wall, then the surface normal at the live cell next to the wall is
\begin{equation}
\hat{n}=  \hat{n} _{ w }  cos  \theta _{ w }  +  \hat{m} _{ w } sin  \theta _{ w } 
\end{equation}

where $\hat{n}$ and $\hat{m}$ are the unit vectors normal and tangential to the wall, respectively \citep{fluent}. This implementation of the surface tension force can cause unphysical velocity near the interface, known as \textit{spurious currents} \citep{harvie2006}, due to unbalanced representation of surface tension induced capillary force and related pressure jump across the interface. There have been some attempts to reduce such phenomena \citep{hirt-1981, popinet1999, meier2002}. In this work, an explicit geometric reconstruction scheme is used to represent the interface by using a piecewise\textendash linear approach in VOF method \citep{youngs1982}. \citet{gupta-2009} observed spurious pressure oscillations using Green\textendash Gauss \textit{cell} based gradient schemes. Consequently, an alternative approach using Green\textendash Gauss \textit{node} based schemes \citep{seifollahi2008} is suggested to accurately calculate the gradients that can help in minimizing the spurious currents. Therefore, in this study, Green\textendash Gauss \textit{node} based gradient calculation is used to avoid such unphysical situations.    

\section{Solution methodology}

\begin{figure}[h]
	\centering
	\includegraphics[width=1\textwidth]{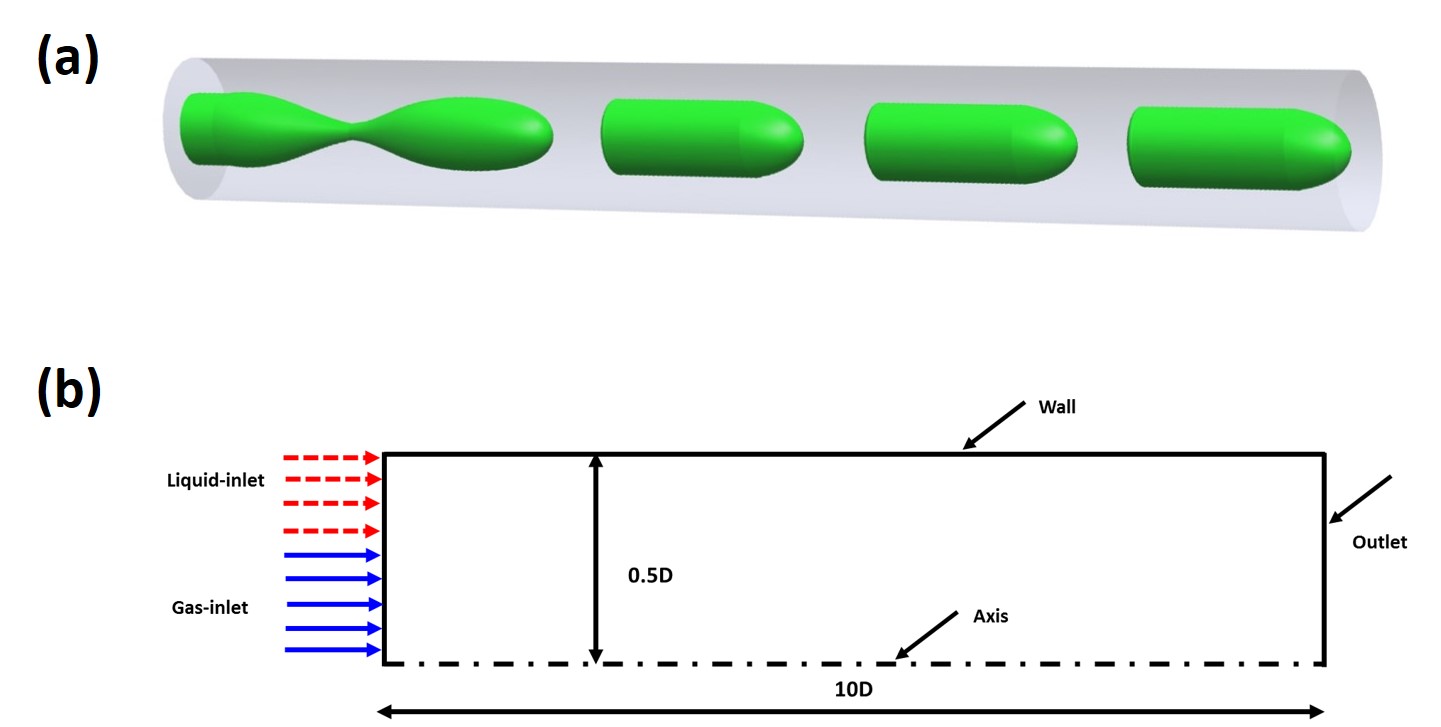}
	\caption{\label{fig:refine_mesh}(a) 3D schematic representation Taylor bubble formation in a circular microchannel, and (b) computational domain with dimension and imposed boundary condition.}
\end{figure}

Schematic of the modeled geometry to understand the hydrodynamic characteristics of gas\textendash liquid two\textendash phase flow in a circular microchannel is shown in Fig.~\ref{fig:refine_mesh}a. A two dimensional axisymmetric geometry is considered for circular microchannel having a diameter (D) of 0.5 mm and a length of 10D, as shown in Fig.~\ref{fig:refine_mesh}b \citep{asadoi2011}. As the liquid film thickness around the Taylor bubble in microchannels is nonuniform, which varies in both lateral and axial directions from the bubble nose to its tail \citep{breth-1961}, the effect of mesh element size on the bubble length is initially investigated for different numbers of mesh elements. Following the guidelines of \citet{gupta-2009} and \citet{Fletcher2016}, quadrilateral mesh elements are utilized for the solution. Grid independence study was performed with four different mesh element size of $2~\mu m$, $3~\mu m$, $4~\mu m$, and $5~\mu m$. Optimum mesh element size is found to be 5 $\mu m$ and the total number of mesh elements are 50,000. The results presented henceforth is based on element size of 5 $\mu m$ in the core region. To capture the sharp interface and the liquid film thickness around the Taylor bubble, mesh elements near the wall are further refined \cite{gupta-2009}.  

Transient simulations are carried out using the pressure based solver in ANSYS Fluent 17.0. Pressure\textendash velocity coupling is approximated by Fractional Step Method (FSM) using first order implicit Non\textendash Iterative Time Advancement (NITA) scheme \citep{fluent}. An implicit body force treatment is used to take into account the surface tension forces. Quadratic upstream interpolation for convective kinetics (QUICK) \cite{leonard1979} and geo\textendash reconstruct \citep{youngs1982} schemes are used for the momentum and volume fraction equation discretization, respectively. Variable time step and fixed Courant number (Co = 0.25) are used for solving momentum and pressure equations. At the liquid and gas inlet, constant velocity boundary condition is imposed and the pressure outlet boundary condition is set for the microchannel outlet. No\textendash slip wall boundary condition is applied to the impermeable wall.


\section{Results and discussion}
\begin{figure}[h]
	\centering
	\includegraphics[width=\textwidth]{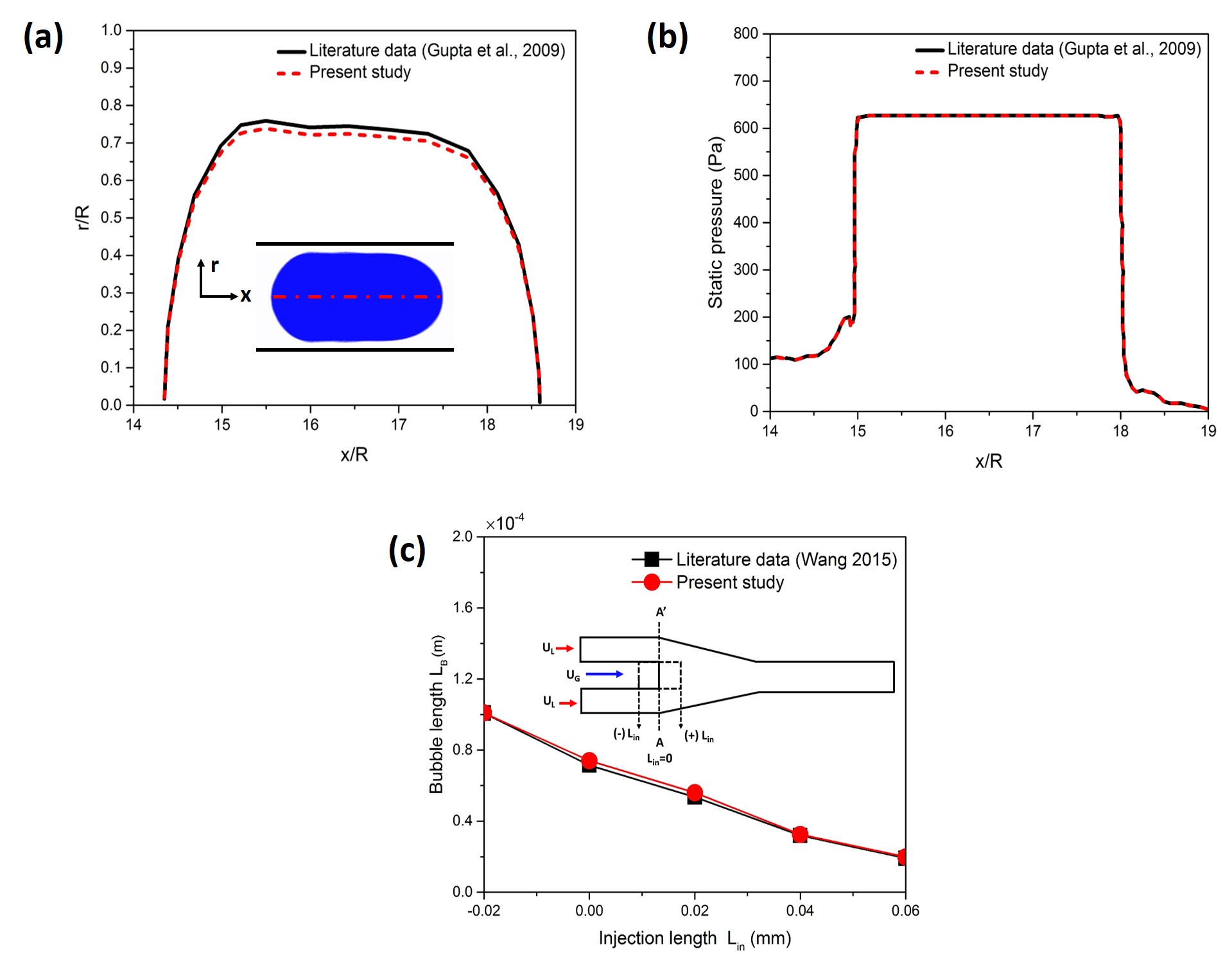}
	\caption{\label{fig:Wang_validation} Comparison of (a) Taylor bubble shape, (b) axial pressure distribution in a unit cell obtained from \citet{gupta-2009} and this work for air-water system with $U_G=0.5~m/s$, $U_L=0.5~m/s$, and $\eta_w$= 8.899$\times10^{-4}$ kg/m.s at 0.0085 s, and (c) Taylor bubble length for different injection length ($L_{in}$) at $\eta_w$= 1.003$\times10^{-3}$ kg/m.s, $Q_{G}= 0.47$ $\mu L/s$, and $Q_{L}= 2.01$ $\mu L/s$ with the results of \citet{wangg-2015}. The inset shows simulated geometry with varying gas injection position.}
\end{figure}

At first, the developed model is verified for bubble shape obtained in a Newtonian (air-water) system\cite{gupta-2009}. The comparison of a steady bubble interface extracted from the results of \citet{gupta-2009} and those obtained from our CFD simulations for $U_G=0.5~m/s$, $U_L=0.5~m/s$, and $\eta_w$= 8.899$\times10^{-4}$ kg/m.s at 0.0085 s is shown in Fig. \ref{fig:Wang_validation}a. Fig. \ref{fig:Wang_validation}b illustrates the comparison of axial pressure distribution on the axis of the channel in a unit cell with the results of \citet{gupta-2009} under identical operating condition. 

The developed model is further used to calculate and corroborate the bubble length in a tapered co-flow geometry, as reported in \citet{wangg-2015}. The inset in Fig. \ref{fig:Wang_validation}c describes the simulated geometry with varying gas phase injection position. A reference line ($AA^{'}$) is marked to identify the onset of tapering section and is assigned as zero injection length ($L_{in}$). In case of negative injection length, gas phase is introduced at a position prior to this reference line ($AA^{'}$). Fig. \ref{fig:Wang_validation}c shows the comparison of bubble lengths reported by \citet{wangg-2015} and those predicted by this work for various gas injection positions. In all cases, comparison of our model predictions shows excellent agreement with the literature data.

Armed with the well validated model, in the following section, we discuss the results of Newtonian liquid (water) and gas (air) phase in a co\textendash flow geometry, as illustrated in Fig. \ref{fig:refine_mesh}a. Subsequently, the liquid phase is considered as aqueous solution of carboxymethyl cellulose (CMC) with different mass concentration to understand the Taylor bubble hydrodynamics in non\textendash Newtonian liquid phase systems. The liquid inlet velocity ($U_L$), and surface tension ($\sigma$) values are ranged from 0.25 to 1.0 m/s, and 0.025 to 0.1 N/m, respectively, for this study. In all cases, simulations are performed for sufficiently longer time than the corresponding residence time to attain a steady state flow behavior and thereafter, the analyses are carried out. 

\subsection{Newtonian liquid phase}
\subsubsection{Effect of surface tension}
\begin{figure}[h]
	\centering
	\includegraphics[width=0.7\textwidth]{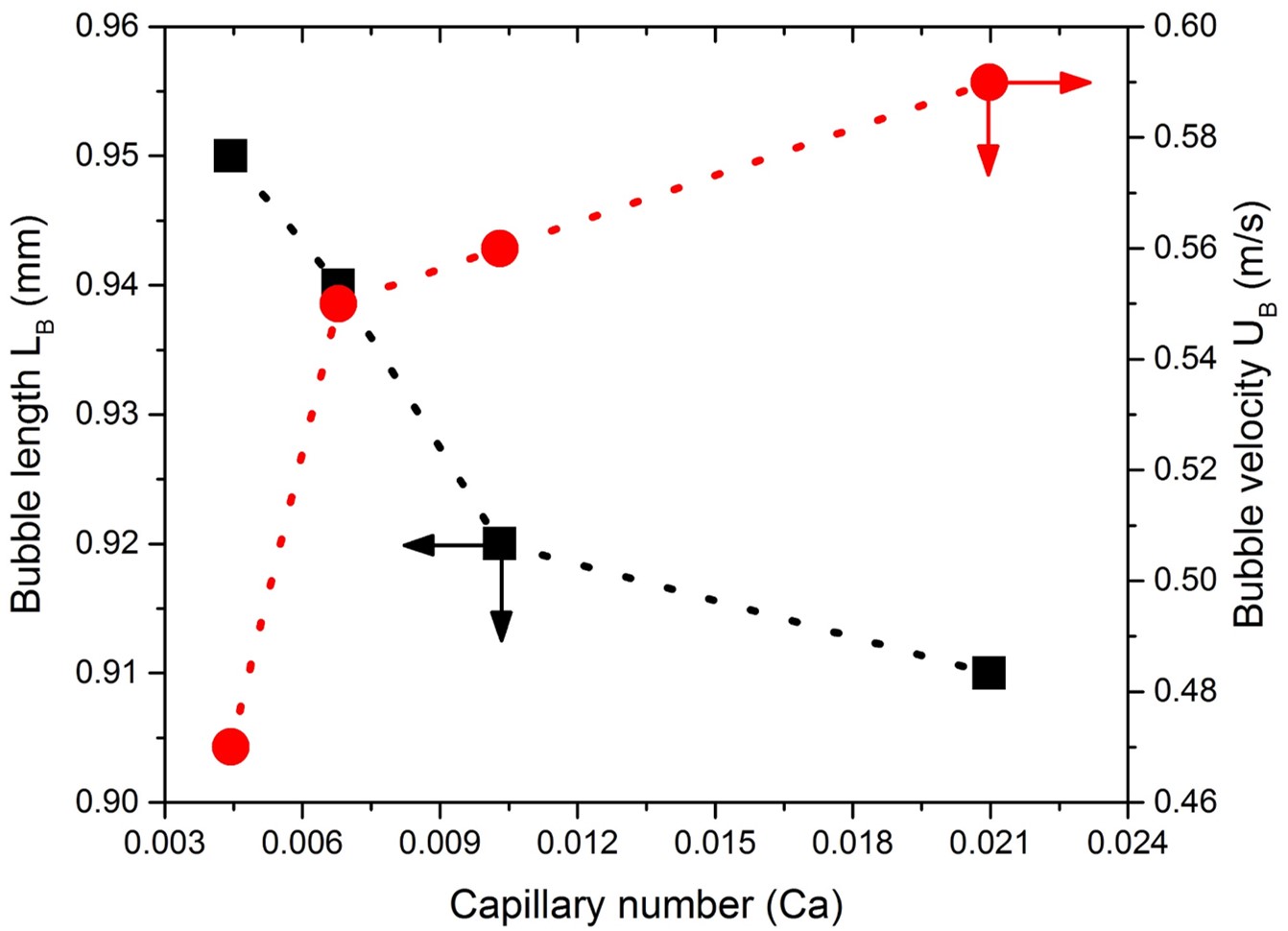}
	\caption{\label{fig:Surfacetension_1} Effect of Capillary number on the bubble length and velocity at fixed operating condition of $\eta _{W}  = 8.899\times10^{-4}$ kg/m.s, $U_{L}$ = 0.5 m/s and $U_{G}$ = 0.5 m/s.}
\end{figure}

Taylor bubble flow in microchannels has a strong dependence on the surface tension and viscous forces. Fig.~\ref{fig:Surfacetension_1} illustrates the effect of surface tension on the bubble length and velocity.
It can be observed that with increasing $Ca$ ($= \frac{\eta_L U_B}{ \sigma}$), Taylor bubble length decreases. This is attributed to the fact that at lower surface tension, weak interfacial force between two phases leads to smaller bubble and higher formation frequency. For all cases, liquid film thicknesses near the wall are calculated and compared with the available correlations that are based on various approaches and/or measurement techniques. To measure the liquid film thickness a vertical reference line is drawn in the middle of bubble and the film thickness is estimated from the gas\textendash liquid interface to the wall as shown in \ref{fig:Taylorbubble_model}b. Table S1 (in Supporting Information) summarizes the list of methods and respective correlations\citep{breth-1961,irand-1989,aussi-2000} that are considered in this work, and the comparison is provided in Table \ref{tab:Surfacetension_correlation}. 

%
%
%
%
%
%

\begin{table}[!h]
	\small
	\centering
	\caption{Comparison of film thickness values with available correlations}
	\label{tab:Surfacetension_correlation}
	\begin{tabular}{lcccc}
		
		\hline
		\multirow{2}{*}{Reference} & \multicolumn{4}{c}{Film thickness ($\delta$)$\mu$m}                                                                                            \\  \cline{2-5}
		& \multicolumn{1}{c}{$Ca=0.0210$} & \multicolumn{1}{c}{$Ca=0.0103$} & \multicolumn{1}{c}{$Ca=0.0068$} & \multicolumn{1}{c}{$Ca=0.0044$} \\ \hline
		\citet{breth-1961}  & 25.48  &  15.87 &  12.01  &  9.06 \\
		\citet{irand-1989}  & 28.59 & 20.64 & 16.89 &  13.72 \\
		\citet{aussi-2000}  & 20.31   & 13.70  & 10.72  & 8.30 \\
		This work (CFD) & 23.90 & 18.91 & 15.04 & 11.20 \\
		\hline
		
	\end{tabular}
\end{table}

For $10^{-3}$ $\leq Ca \leq$ $10^{-2}$, our results conform arguably well with the calculations using lubrication theory \citep{breth-1961}. From Table \ref{tab:Surfacetension_correlation}, it is also apparent that with increasing $Ca$, liquid film thickness near the wall increases, which in turn enhances the bubble velocity, as shown in Fig. \ref{fig:Surfacetension_1}. This is ascribed to lower resistance between the bubble and solid wall with increasing film thickness. Such phenomena is well supported by the lubrication theory for low $Ca$ values\citep{breth-1961, klas-2014}. In all cases, our predicted results closely match with \citet{breth-1961} correlation. 

\begin{figure}[h]
	\centering
	\includegraphics[width=0.7\textwidth]{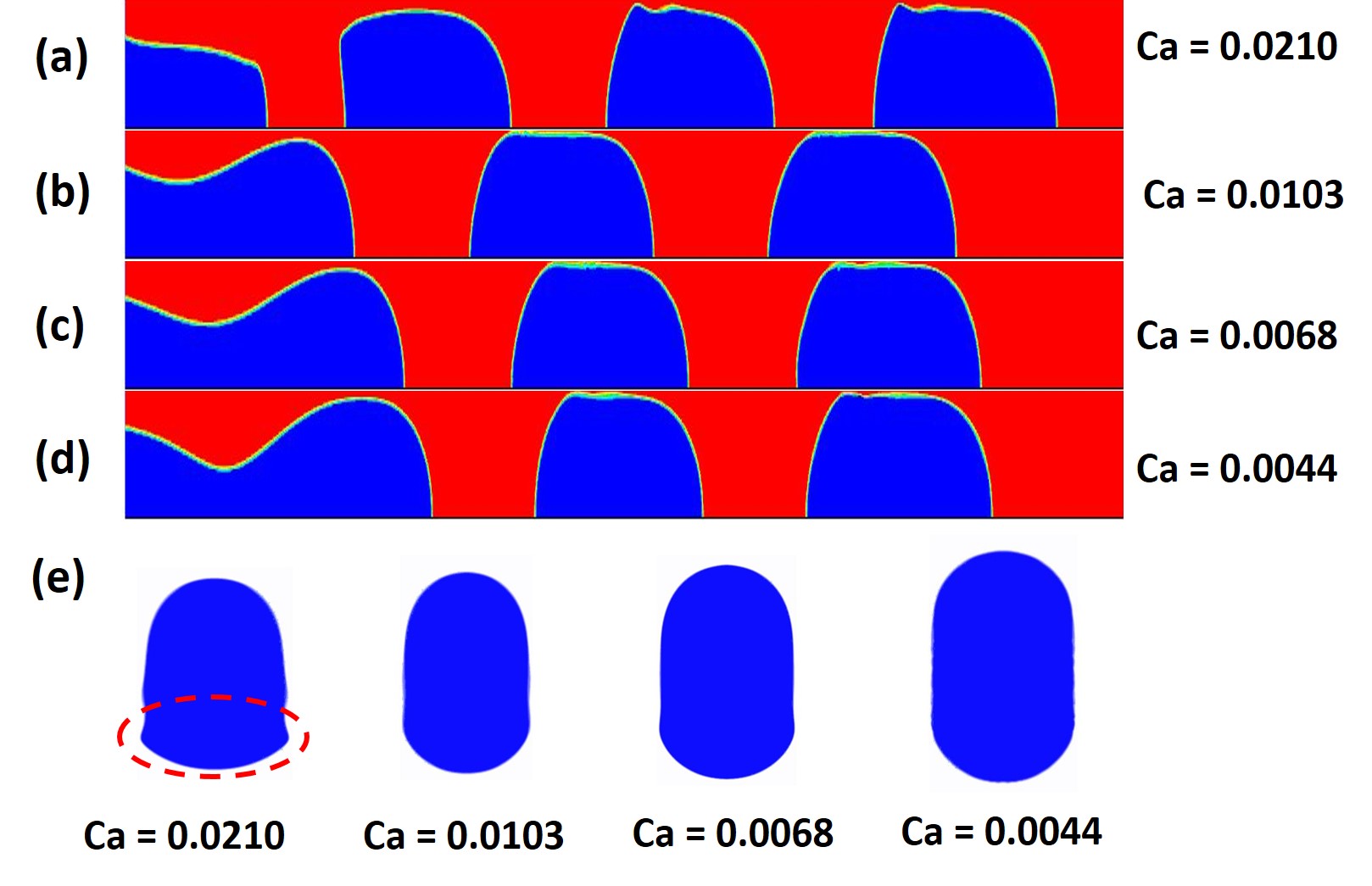}
	\caption{\label{fig:Surfacetension_2}  Phase volume fraction contours (color blue: gas, color red: liquid) at 8 ms for $Ca=$(a) 0.0210, (b) 0.0103, (c) 0.0068, (d) 0.0044, and (e) Taylor bubble shape for various surface tension values at $\eta _{W} = 8.899\times10^{-4}$ kg/m.s, $U_{L}$ = 0.5 m/s and $U_{G}$ = 0.5 m/s.}
\end{figure}

From the phase volume fraction contours (Fig. \ref{fig:Surfacetension_2} a\textendash d), it can be noticed that bubble detachment occurs rapidly at lower surface tension values. Therefore, for constant mass flow rates, the number of bubbles increases with decreasing surface tension. For higher surface tension i.e., at low $Ca$ values, the front and rear curvatures of the bubble are found to be similar (Fig. \ref{fig:Surfacetension_2}e). However, at higher $Ca$, bubble nose (refer to Zone: III in Fig.~\ref{fig:Taylorbubble_model}a) becomes sharper and bullet-shaped with some wiggles at the back due to weak interfacial tension between phases.

\begin{figure}[h]
	\centering
	\includegraphics[width=0.7\textwidth]{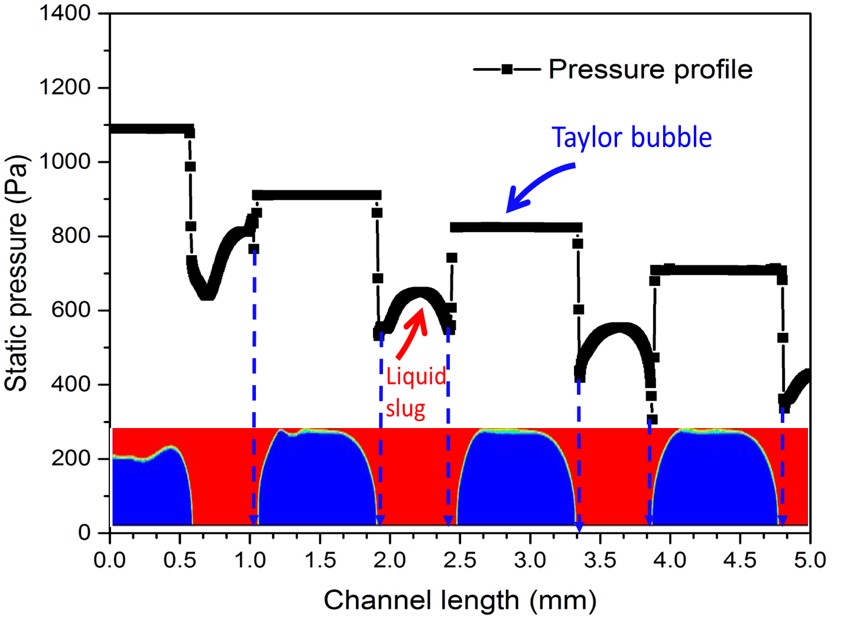}
	\caption{\label{fig:Surfacetension_3} Pressure distribution along the channel axis length for a fixed operating condition with $Ca = 0.0061$, $U_{L}$ = 0.5 m/s and $U_{G}$ = 0.5 m/s.}
\end{figure} 
The pressure difference between the gas bubble and liquid slug combined with dynamic head lead to a pressure field inside the microchannel as shown in Fig.~\ref{fig:Surfacetension_3}.  Due to the differences in curvatures at the nose and rear of the bubble, there also exists a small pressure drop, as shown in Fig. \ref{fig:Surfacetension_3}. This is in line with the observation by other researchers \citep{kreut-2005,cheru-2015}, as well. Moreover, it is evident from Fig. \ref{fig:Surfacetension_3}, that for fixed gas and liquid properties, the total pressure inside the bubble is constant.

\subsubsection{Effect of liquid inlet velocity}
\begin{figure}[h]
	\centering
	\includegraphics[width= 0.7\textwidth]{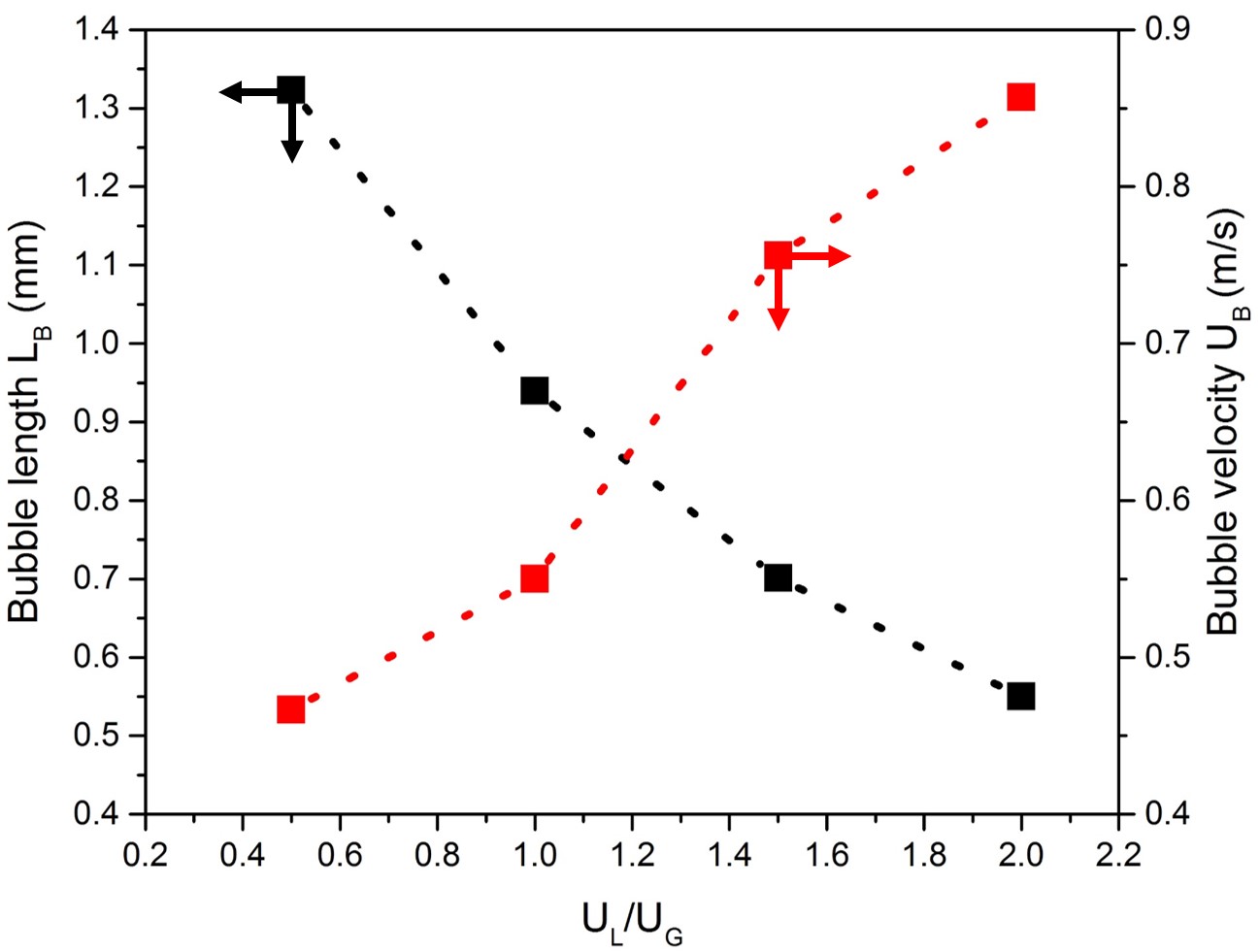}
	\caption{\label{fig:Velocity_1}  Effect of mixture inlet velocity on the bubble length, and bubble velocity at $\sigma$ = 0.072 N/m, and $\eta _{W}  = 8.899\times10^{-4}$ kg/m.s.}
\end{figure}

\begin{figure}[h]
 	\centering
 	\includegraphics[width=0.7\textwidth]{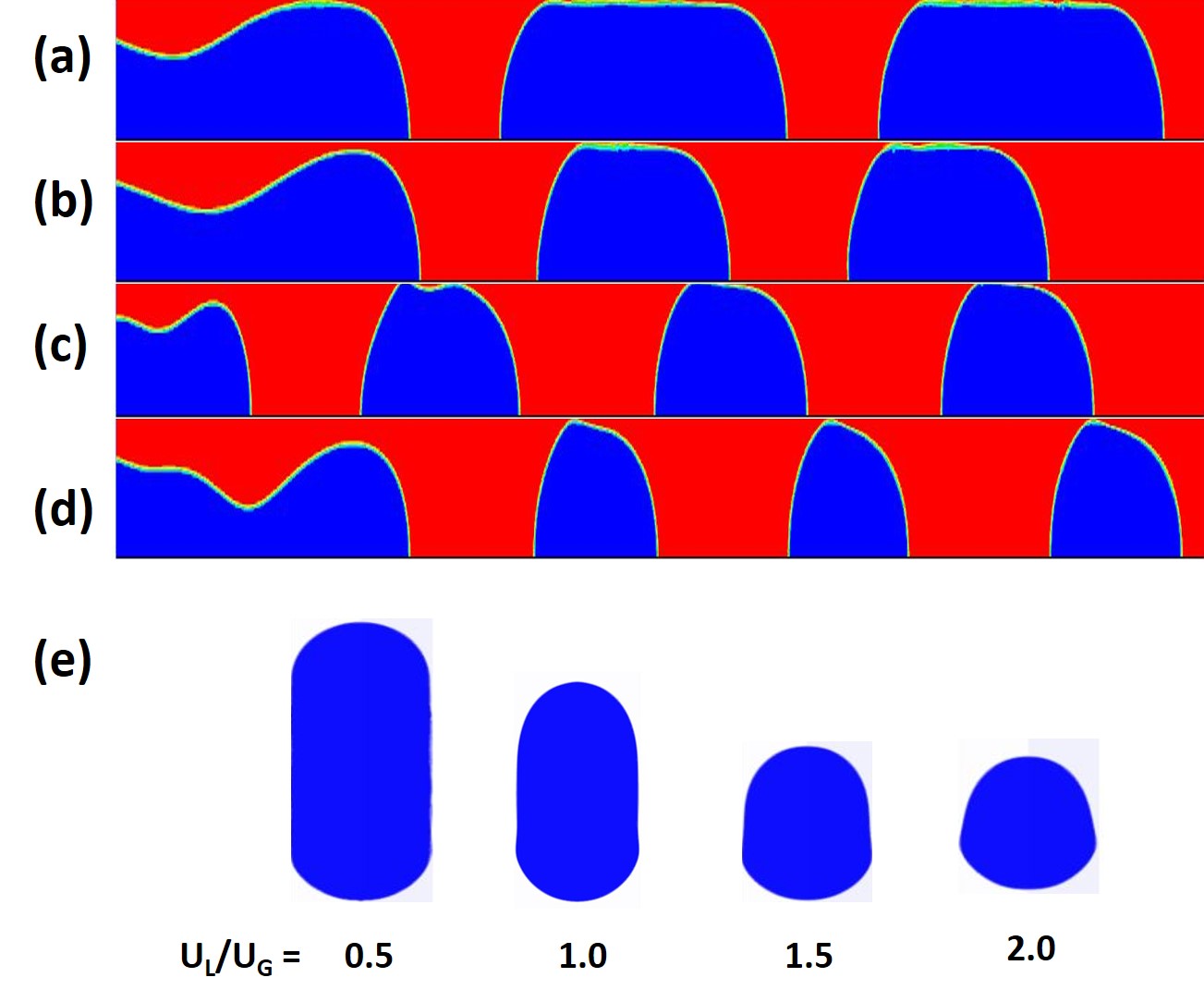}
 	\caption{\label{fig:Velocity_2} Phase volume fraction contours (color blue: gas, color red: liquid) for $U_L/U_G$ = (a) 0.5 at 0.018 s, (b) 1.0 at 0.011 s, (c) 1.50 at 0.008 s, (d) 2.0 at 0.006 s, and (e) corresponding steady bubble shapes for various velocity ratios at $U_G$ = 0.50 m/s, $\sigma$ = 0.072 N/m, and $\eta _{W} = 8.899\times10^{-4}$ kg/m.s.}
 \end{figure}

The effect of liquid phase inlet velocity ($U_L$) on Taylor bubble formation is observed by keeping gas phase velocity ($U_G$) and other fluid properties are at constant. Fig. \ref{fig:Velocity_1} shows that with increasing $U_L/U_G$ ratio, bubble length decreases.  
This is a consequence of increased shear force on the gas phase that in turn also results in change of bubble shape at higher velocity ratios. Fig. \ref{fig:Velocity_2} qualitatively describes this effect on the Taylor bubble formation phenomena. In particular, Fig. \ref{fig:Velocity_2}e helps to clearly understand the effect of liquid inlet velocity on the shape of Taylor bubble. Bubble shapes are found to be similar at lower $U_L/U_G$ values. However, the shape changes significantly at high $U_L/U_G$ (=2.0) primarily due to considerable enhancement in inertia force as shown in Fig. \ref{fig:Velocity_2}e. 


In addition to bubble length observation, bubble velocity and liquid film thickness are also analyzed. The bubble velocity is found to rapidly increase with the enhanced liquid inlet velocity (Fig. \ref{fig:Velocity_1}) and the surrounding liquid film thickness also increases, as mentioned in Table \ref{tab:velocity_correlation}.
\begin{table}[!h]
	\centering
	\small
	\caption{Comparison  of film thickness between available correlation with numerical results  }
	\label{tab:velocity_correlation}
	\begin{tabular}{lcccc}
		\hline
		\multirow{2}{*}{Reference} & \multicolumn{4}{c}{Film thickness ($\delta$)$\mu$m}                                                                                            \\  
		
		\cline{2-5}
		& \multicolumn{1}{c}{$Ca=0.0058$} & \multicolumn{1}{c}{$Ca=0.0068$} & \multicolumn{1}{c}{$Ca=0.0093$} & \multicolumn{1}{c}{$Ca=0.0106$} \\ \hline
		
		\citet{breth-1961}  & 10.77  &  12.01 &  14.85  &  16.15 \\
		\citet{irand-1989}  & 15.59 & 16.89 & 19.69 &  20.89 \\
		\citet{aussi-2000}  & 9.72 & 10.72 & 12.93  & 13.90 \\
		This work (CFD) & 12.93 & 15.04 & 17.33 & 19.26 \\
		\hline
		
	\end{tabular}
\end{table}

Our results are in reasonable agreement with the film thickness values estimated from the correlation based on lubrication theory \citep{breth-1961} at moderate $Ca$ values, and are in excellent agreement with the analytical results of \citet{irand-1989}.

\subsection{Non\textendash Newtonian liquid phase}

\subsubsection{Effect of CMC concentration}
\begin{figure}[h]
	\centering
	\includegraphics[width=1\textwidth]{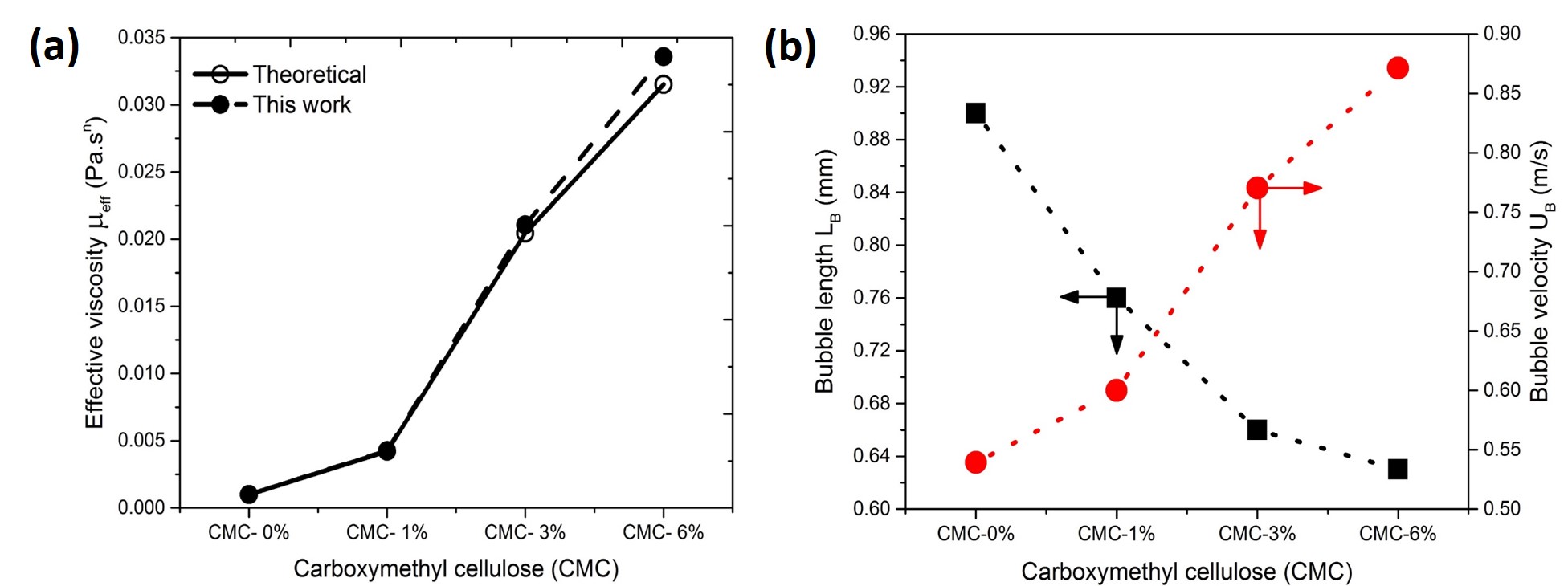}
	\caption{\label{fig:Power_law_CMC1} Effect of CMC concentration on (a) effective viscosity, (b) Taylor bubble length and bubble velocity at $U_{L}$ = 0.5 m/s and $U_{G}$ = 0.5 m/s.}
\end{figure}

To understand the Taylor bubble behavior in non-Newtonian liquid phase, we have considered different mass concentration of carboxymethyl cellulose (CMC) in water as the continuous liquid phase that exhibit power\textendash law behavior (Eq. \ref{eq:nnvis_eqn}). Properties of three different aqueous solutions of CMC are taken from the experimental results of \citet{picchi-2015} that are found to be shear thinning ($n$ \textless 1) in nature. These rheological properties are listed in Table S2 (in Supporting Information). It can be noted that the continuous phase becomes Newtonian (water) when CMC concentration is zero. This systematic selection of data helps us to identify the influence of non-Newtonian continuous phase and to compare model predictions with Newtonian results. 
To understand the effect of CMC concentration in terms of bulk viscosity and overall viscous force, effective viscosities ($\eta _{eff}$) of various solutions are estimated from the simulations and compared with the results derived from Eq. \ref{eq:effective_eqn}\citep{rhodes-2008}.

\begin{equation}
	\label{eq:effective_eqn}
	\eta_{eff}=k\left(\frac{3n+1}{4n}\right)^{n}\left(\frac{8 U_{L} }{D}\right)^{n-1}
\end{equation}
where $K$, $U_L$, $D$, and $n$ are consistency index, liquid inlet velocity, diameter of the channel, and power\textendash law index, respectively. Fig. \ref{fig:Power_law_CMC1}a shows enhanced effective viscosity with increasing CMC concentration and the CFD calculations are in excellent agreement with the theoretical values.

As a result of higher viscous force, bubble length is observed to decrease with increasing CMC concentration (Fig. \ref{fig:Power_law_CMC1}b). It can also be observed from Fig. \ref{fig:Power_law_CMC1}b that bubble velocity also changes significantly.  
Similar to the discussions in earlier sections, this is attributed to the influence of surrounding liquid film thickness that is evaluated and provided in Table~\ref{tab:non_newtonian_correlation}. It shows that modified Capillary number ($Ca^*=KU_B^nD^{(1-n)}/\sigma$) increases with increasing CMC concentration for a fixed set of inlet velocities. Consequently, liquid film around the Taylor bubble becomes thicker. Interestingly, in this case, film thickness results are very close to the estimated values from the correlation proposed by \citet{aussi-2000} through experimental observations in Newtonian liquids. 

\begin{table}[!h]
	\centering
	\caption{Comparison of non\textendash Newtonian liquid film thickness with available correlations}
	\label{tab:non_newtonian_correlation}
	\begin{tabular}{lcccc}
		\hline
		\multirow{2}{*}{Reference} & \multicolumn{4}{c}{Film thickness ($\delta$)$\mu$m}                                                                                            \\  \cline{2-5}
		& \multicolumn{1}{c}{CMC\textendash0\%} & \multicolumn{1}{c}{CMC\textendash1\%} & \multicolumn{1}{c}{CMC\textendash3\%} & \multicolumn{1}{c}{CMC\textendash6\%} \\
		\vspace{-0.5cm}
		\multirow{2}{*}{} & \multicolumn{4}{c}{}                                                                                            \\  \cline{2-5}
		& \multicolumn{1}{c}{$Ca^*=0.0076$} & \multicolumn{1}{c}{$Ca^*=0.0387$} & \multicolumn{1}{c}{$Ca^*=0.2572$} & \multicolumn{1}{c}{$Ca^*=0.4936$} \\
		\hline
		\citet{breth-1961} & 12.95 &	38.31 & 135.49 & 209.25\\
		\citet{irand-1989} & 17.83 &	37.12 &	69.50 &	79.02 \\
		\citet{aussi-2000} & 11.47 &	27.70 &	57.54 &	67.66 \\
		This work (CFD) & 16.41 & 26.51 & 55.92 & 64.21 \\
		\hline
		Formation frequency $f$(kHz) & 0.40 & 0.50 & 0.80 & 0.87 \\
		\hline	
	\end{tabular}
\end{table}

\begin{figure}[!h]
	\centering
	\includegraphics[width=0.7\textwidth]{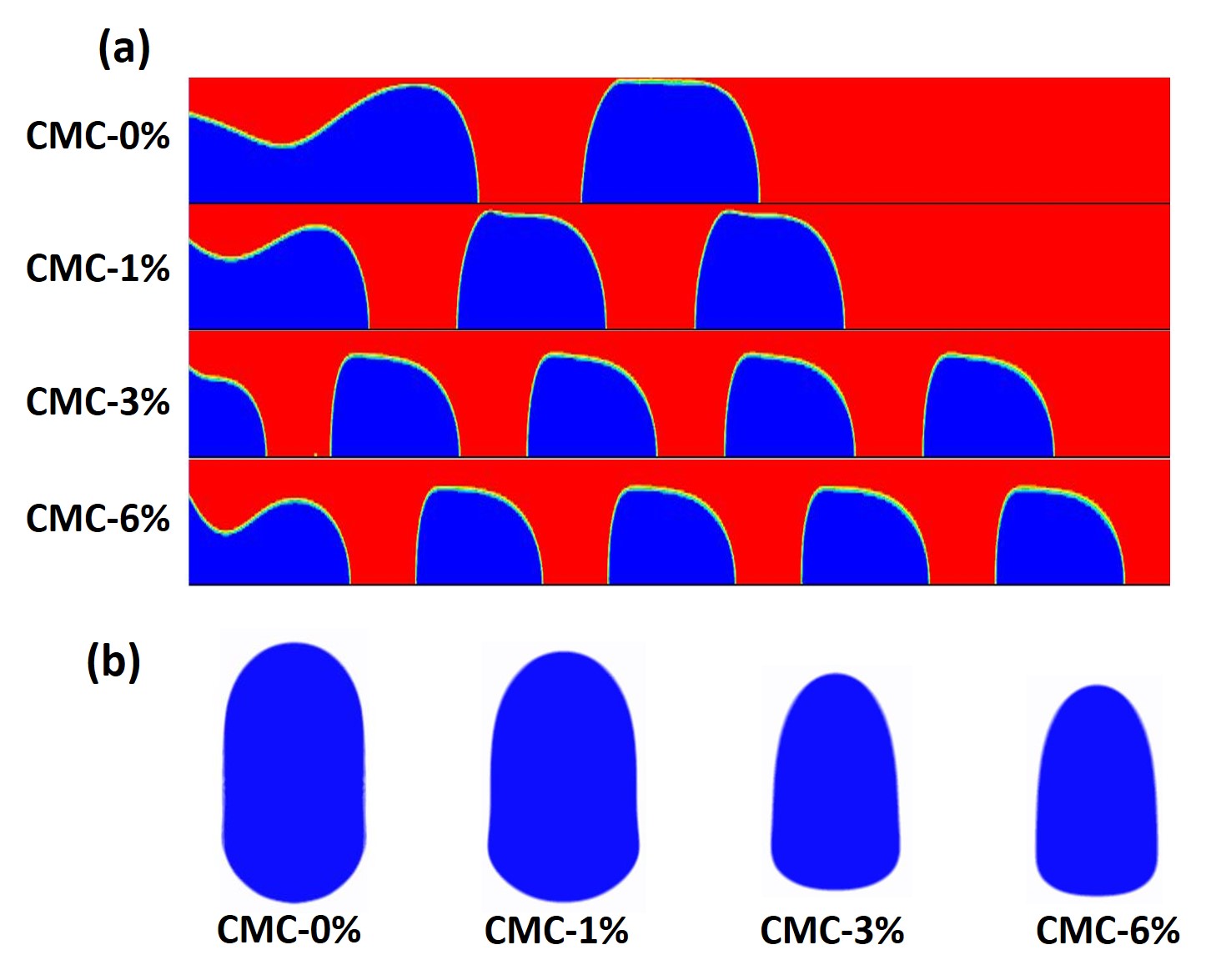}
	\caption{\label{fig:Power_law_CMC2} (a) Phase volume fraction contours (color blue: gas, color red: liquid) at 5.5 ms for different CMC concentration, (b) Taylor bubble shape in various of CMC concentration for $U_{L}$ = 0.5 m/s and $U_{G}$ = 0.5 m/s.}
\end{figure}

Fig.~\ref{fig:Power_law_CMC2}a shows that increasing CMC concentration leads to more frequent bubble formation for constant inlet velocities. 
This is also quantitatively mentioned in Table \ref{tab:non_newtonian_correlation}. The rheological properties of the continuous phase are found to influence the nose curvature of the Taylor bubble, which is mainly controlled by the balance between capillary and inertial forces\citep{breth-1961}. It is evident from Fig.~\ref{fig:Power_law_CMC2}b that higher $\eta _{eff}$ imparts significant alteration on bubble shape in the dynamic meniscus and nose region (Zone: II and Zone: III, respectively in Fig.~\ref{fig:Taylorbubble_model}). 

Fig.~\ref{fig:Power_law_CMC2}b also demonstrates that with increasing CMC concentration, the radius of curvature ($R$ in Fig.~\ref{fig:Taylorbubble_model}b) decreases but the length of nose increases, and the bubble tends to be bullet shaped. This transition arises while increasing the CMC concentration and particularly in situations where viscous force of the continuous phase overcomes the interfacial force. To understand the influence of shear thinning behavior, velocity profiles in liquid slug and gas bubble are also analyzed for two extreme cases (CMC\textendash0\% i.e., water, and CMC\textendash6\% solution).

 \begin{figure}[h]
	\centering
	\includegraphics[width=0.9\textwidth]{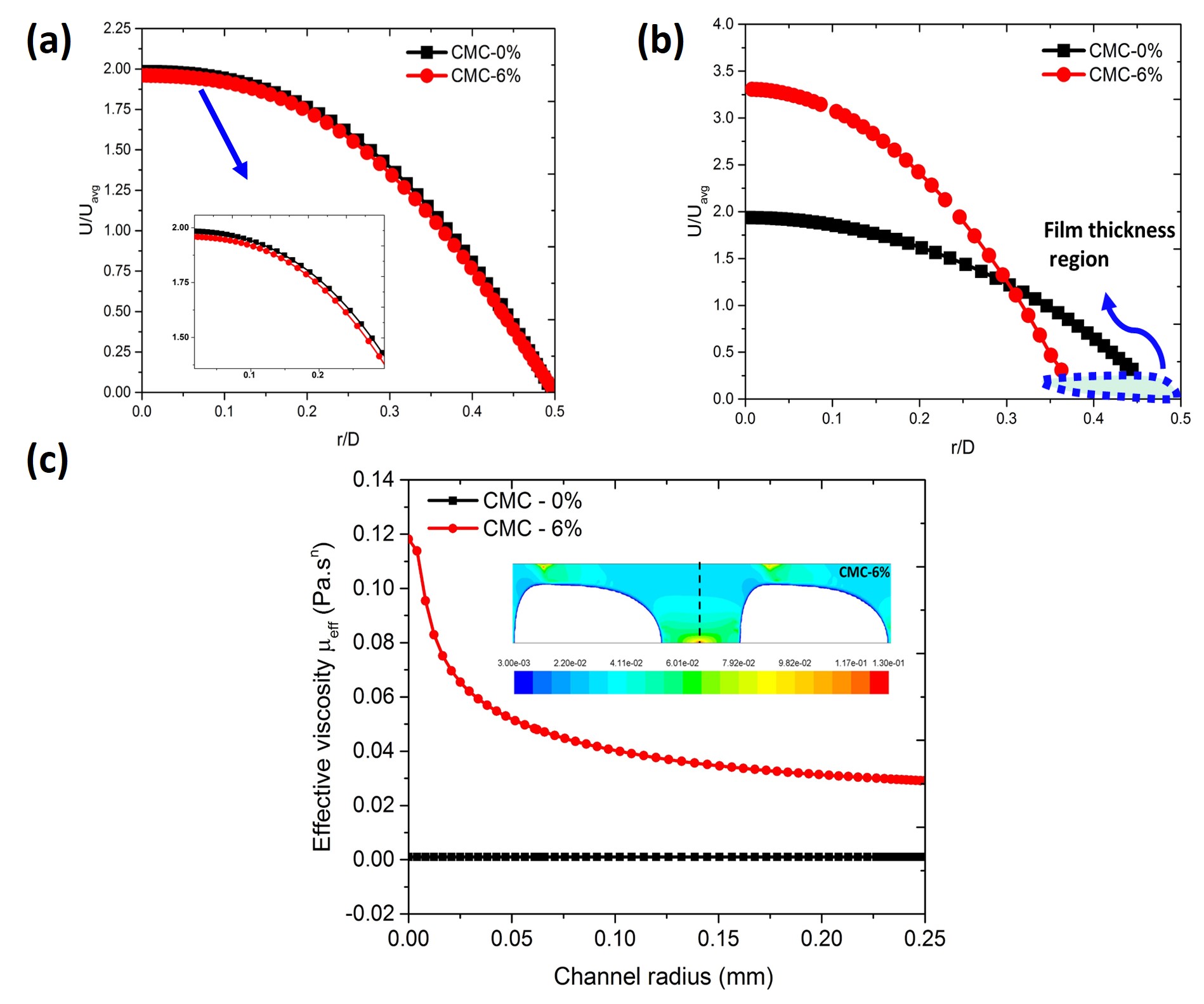}
	\caption{\label{fig:Power_law_CMC1_1} Effect of CMC concentration on velocity profiles in the middle of (a) slug, and (b) bubble at $U_{L}$ = 0.5 m/s and $U_{G}$ = 0.5 m/s. (c) Distribution of effective viscosity in the middle of liquid slug at $U_{L}$ = 0.5 m/s and $U_{G}$ = 0.5 m/s. The inset shows contour of non\textendash homogeneous viscosity distribution of CMC\textendash6\%  where the middle of liquid slug is marked by dashed line.}
\end{figure}

Figs.~\ref{fig:Power_law_CMC1_1}a and \ref{fig:Power_law_CMC1_1}b show non-dimensional velocity profiles at the middle of the slug and bubble, respectively. In case of Newtonian system, parabolic profile is evident in the slug indicating fully developed laminar flow however relatively flatter profile is evident in non-Newtonian system which is the typical characteristic of laminar velocity profile of a shear thinning fluid. However, considerably higher velocity is observed within the bubble for non-Newtonian system. This is anticipated due to enhanced film thickness and bullet shaped bubble \cite{angeli2017}. The non\textendash homogeneous viscosity profile, presented in Fig.~\ref{fig:Power_law_CMC1_1}c (inset), portray the influence of shear thinning CMC solution on the hydrodynamics of Taylor bubble. The distribution of effective viscosity in the middle of the liquid slug for CMC-6\% against the Newtonian liquid i.e., CMC-0\% is also apparent in Fig. ~\ref{fig:Power_law_CMC1_1}c.

 
\subsubsection{Effect of liquid phase velocity}
\begin{figure}[h]
	\centering
	\includegraphics[width=\textwidth]{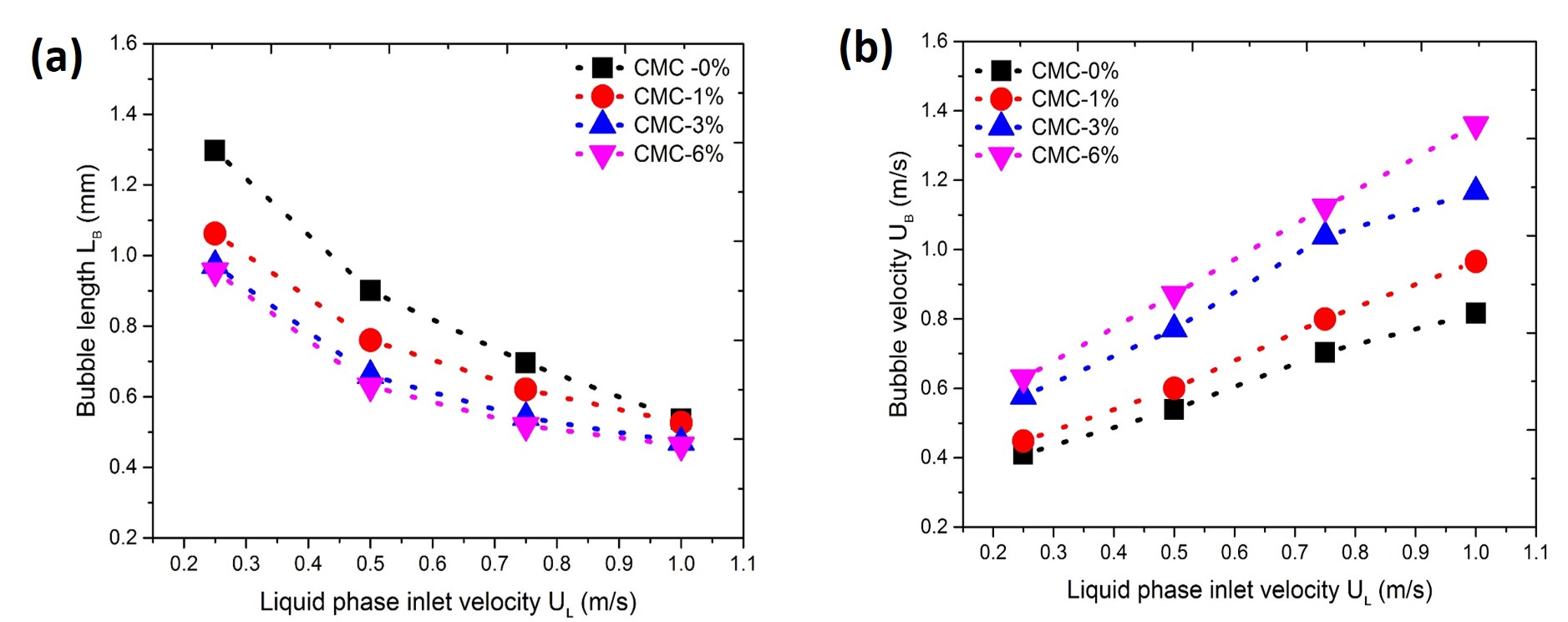}
	\caption{\label{fig:Power_law_CMC3} Effect of liquid phase inlet velocity on bubble (a) length, and (b) velocity for Newtonian and three aqueous CMC solutions at $U_{L}$ = 0.5 m/s and $U_{G}$ = 0.5 m/s.}
\end{figure}

The effect of non-Newtonian liquid inlet velocity on the bubble length and shape is systematically investigated with three CMC concentrations and compared with the Newtonian liquid (CMC-0\%). With increasing liquid phase inlet velocity, similar to the observations in case of Newtonian liquid, bubble length is found to decrease, as shown in Fig.~\ref{fig:Power_law_CMC3}a.

\begin{figure}[ht]
	\centering
	\includegraphics[width=0.8\textwidth]{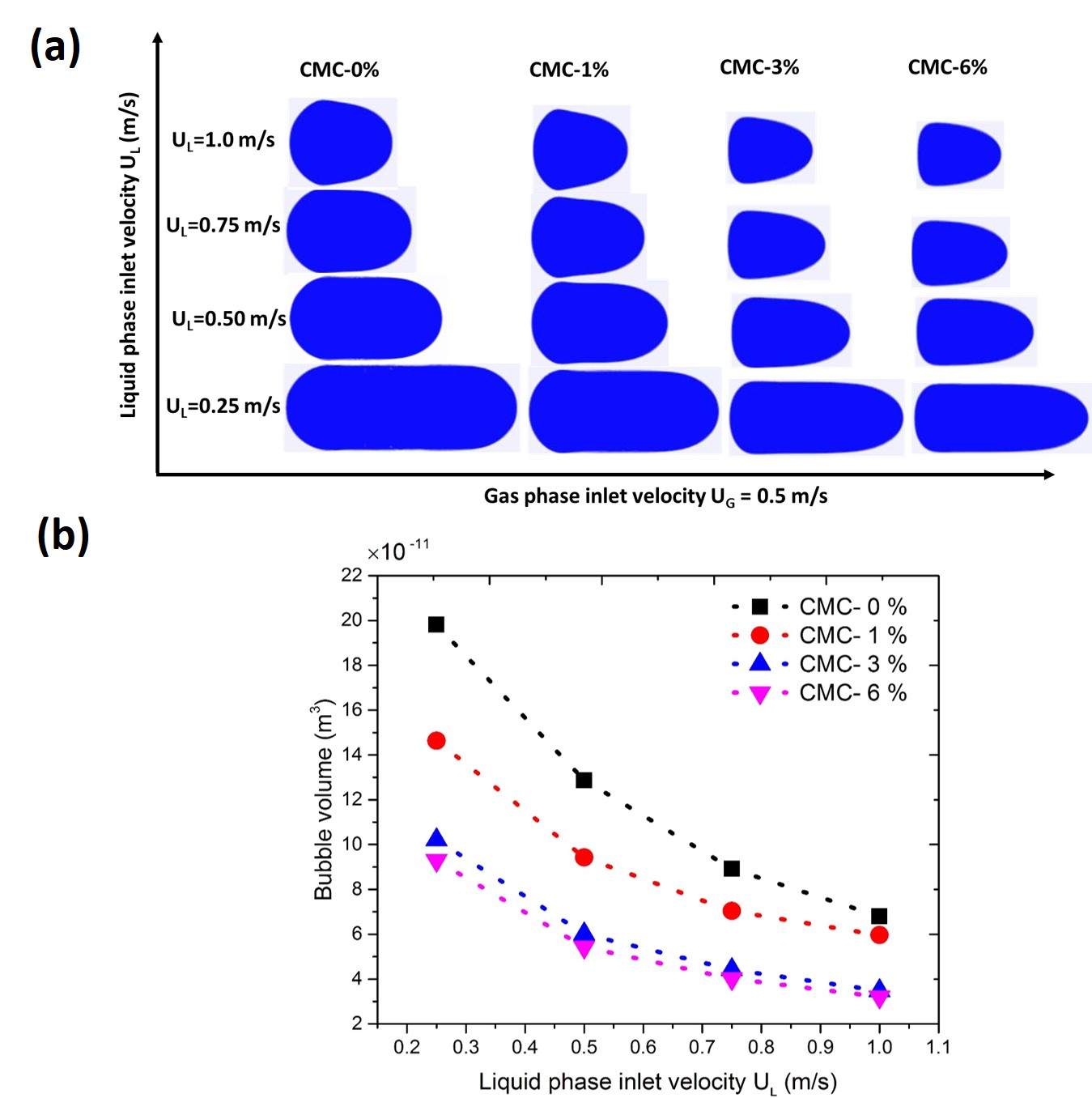}
	\caption{\label{fig:Power_law_CMC4} Effect of liquid phase inlet velocity on Taylor bubble (a) shape, and (b) volume for different CMC concentration.}
\end{figure}

Moreover, it can also be seen from Fig.~\ref{fig:Power_law_CMC3}a that for any particular velocity, bubble length decreases with increasing concentration of CMC. At higher liquid phase velocity, the bubble length is found to be invariant for both Newtonian (CMC-0\%) and non-Newtonian liquids because of the shear-thinning nature of CMC solutions. However, substantial change in the bubble shape is realized, as illustrated in Fig.~\ref{fig:Power_law_CMC4}a. 

\begin{figure}[!ht]
	\centering
	\includegraphics[width=\textwidth]{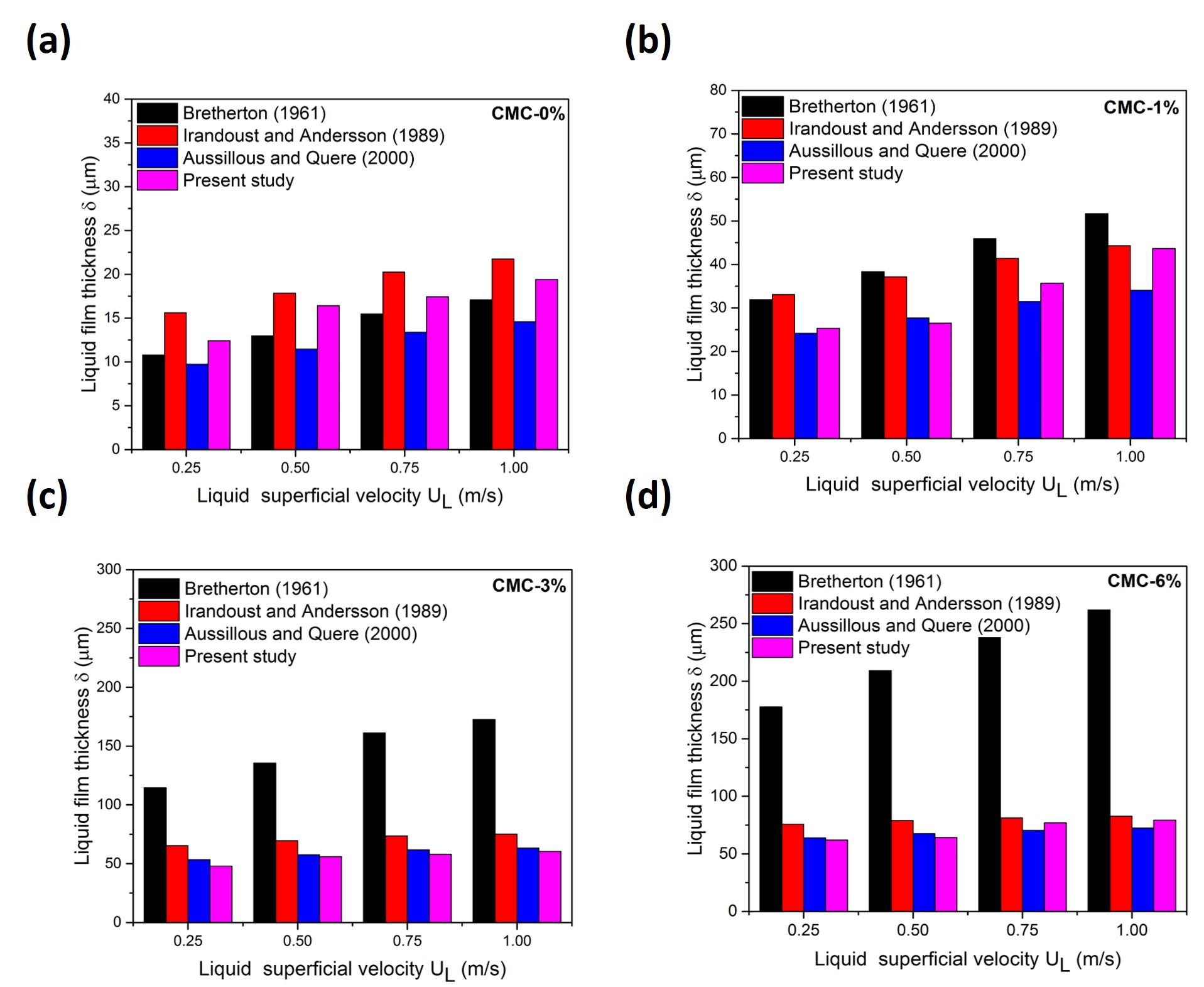}
	\caption{\label{fig:Power_law_CMC5} Comparison of liquid film thickness ($\delta$) predictions against available correlations (a) CMC\textendash0\% (Water), (b) CMC\textendash1\%, (c) CMC\textendash3\%, and (d) CMC\textendash6\%.}
\end{figure}

With increasing CMC concentration and liquid velocity, the shape is found to be bullet shaped. At lower inlet velocities, formation of elongated bubbles are observed in all cases. The Taylor bubble volume is also analyzed and presented in Fig.~\ref{fig:Power_law_CMC4}b that shows reduction in bubble volume with increasing liquid phase velocity for both Newtonian and non\textendash Newtonian cases. It is interesting to note that Taylor bubble volume and shape are almost identical for CMC\textendash 3\% and CMC\textendash 6\% at higher liquid inlet velocity. With increasing liquid velocity, bubble velocity increases and this effect is more pronounced in cases of higher CMC concentration (Fig. \ref{fig:Power_law_CMC3}b). To understand the underlying physics, liquid film thickness is also calculated in these cases and is presented in Fig.~\ref{fig:Power_law_CMC5}. Likewise in Newtonian case, it can be understood that liquid film thickness increases with increasing liquid inlet velocity for any CMC concentration which result in faster movement of the bubble. In cases of higher CMC concentrations, the increase in liquid film thickness is further enhanced due to larger deviation in effective viscosity with increasing liquid superficial velocity.          
It is interesting to note from the comparison of film thickness predictions with available correlations that our results are in well agreement with the proposed correlations of \citet{irand-1989} and \citet{aussi-2000} (Fig. \ref{fig:Power_law_CMC5}). This can be attributed to the applicability of those correlations for a wide range of Ca which encloses the range of our study. However, it is noteworthy that for non-Newtonian liquids, Ca is replaced with $Ca^*$. Specifically, in cases of CMC-3\% and CMC-6\% solutions, the values of $Ca^*$ in our study are beyond the applicable range ($>10^{-2}$) of \citet{breth-1961} correlation. Nonetheless, in cases of Newtonian and lower CMC concentration, i.e. CMC-0\% and CMC-1\%, respectively, as the $Ca^*$ values are within the range of $10^{-3}$ $\leq$ Ca $\leq$ $10^{-2}$, our film thickness results also correspond to the estimation from \citet{breth-1961} correlation.       

\section{Conclusions}
A computational study on the Taylor bubble in Newtonian and non\textendash Newtonian continuous liquid phase with air, as gaseous phase, is carried out in a circular co\textendash flow microchannel. The Taylor bubble length, shape, and velocity are discussed and influencing properties, such as surface tension, inlet velocity, and effective viscosity are reported. Liquid film thickness between the bubble and channel wall is precisely captured to understand its effect on the bubble. For Newtonian liquid, Taylor bubble length is found to decrease with increasing $Ca$ and $U_L/U_G$ ratio. However, the bubble velocity and the liquid film thickness increases. At higher $Ca$ and $U_L/U_G$ ratio, bubble shape changes significantly and wiggles are observed at the rear. To the best of our knowledge, for the first time, the effect of non\textendash Newtonian liquid on the Taylor bubble is computationally investigated using a power\textendash law liquid in a co\textendash flow geometry. Aqueous solutions of CMC with different concentrations, that exhibit shear-thinning behavior, are considered as the non-Newtonian liquid. Effects of CMC concentration and inlet velocity on the Taylor bubble length, shape, and velocity are elaborated and compared with the Newtonian results. Bubble length is found to decrease with increasing concentration of CMC, due to increase in solution effective viscosity. Moreover, the bubble shape changes toward bullet shape with increasing CMC concentration. With increasing liquid inlet velocity, bubble length also decreases but at higher velocity it barely changes with concentration.    

In all cases, liquid film thickness predictions are compared with the correlations proposed for mostly Newtonian liquid. Our results conform arguably well with most of them, and are in excellent agreement specifically, in cases where the values of $Ca$ (for Newtonian) and $Ca^*$ (for non-Newtonian) are within their proposed range of applicability. In particular, for higher concentration of CMC, the results are close to estimated values from \citet{irand-1989} and \citet{aussi-2000}. It can be underlined that the film thickness can be estimated from these correlations within the limit of respective $Ca$ range even in case of non-Newtonian power\textendash law liquid.    


%

\section*{Acknowledgment}
This work is supported by Sponsored Research \& Industrial Consultancy (SRIC), IIT Kharagpur under the scheme for ISIRD (Code: FCF). 

\section*{Nomenclature}


	$Ca$  = Capillary number ($\frac{   \mu_{L} U_B }{\sigma } $) \\
 	$Ca^*$ = modified Capillary number ($KU_B^nD^{(1-n)}/\sigma$) \\
	$D$ = diameter of the channel (m)\\
	$K$ = consistency index ($Pa.s^{n}$) \\
	$L$  = length (m) \\
	$\hat{n}$	= unit normal vector\\
	$p$ = pressure (Pa)\\
	$Q$  = flow rate ($\mu$L/s) \\
	$Re$  = Reynolds number ($\frac{\rho D U_L  }{\eta_{L}} $)\\
	\textit{Greek symbol}\\	
	$\alpha$  = volume fraction\\
	$\dot{\gamma } $ = shear rate (1/s)\\
	$\delta$ = liquid film thickness (m) \\ 
	$\eta$ = dynamic viscosity (kg/m.s)\\
	$\rho$  = density ($kg/m^{3}$)\\
	$\sigma$  = surface tension (N/m)\\ 
	$\tau$ = shear stress (Pa)\\
	\textit{Subscripts}\\
	$B$ = bubble \\
	$G$ = gas\\
	$L$ = liquid

\begin{suppinfo}
Details of rheological properties used in simulations are supplied as Supporting Information. 	
\end{suppinfo}
\bibliography{mybibfile}

\end{document}